\documentclass[12pt]{emulateapj}
\shorttitle{{\it XMM-Newton} observation of PHL~1092}
\shortauthors{Dasgupta, Rao \& Dewangan}
\usepackage{graphicx}
\usepackage{here}
\journalinfo{The Astrophysical Journal}

\begin{document}

\title{{\it XMM-Newton} observation of the narrow-line QSO PHL~1092:
detection of a high and variable soft component }

\author{Surajit Dasgupta\altaffilmark{1},%
A. R. Rao\altaffilmark{1},%
G. C. Dewangan\altaffilmark{2}}
\altaffiltext{1}{%
Department of Astronomy and Astrophysics, Tata Institute of
Fundamental Research, Mumbai-400005, India, email: surajit@tifr.res.in,
arrao@tifr.res.in
}%
\altaffiltext{2}{%
Department of Physics, Carnegie Mellon University, 5000 Forbes
Avenue, Pittsburgh, PA 15213 USA, email: gulabd@cmu.edu
}%


\begin{abstract}
  We present results based on an XMM-Newton observation of the high luminosity
  narrow-line QSO PHL~1092 performed in 2003 January. The $0.3-10$ keV spectrum
  is well described by a model which includes a power-law ($\Gamma\sim2.1$) and
  two blackbody components ($kT \sim 130$ eV and $kT \sim 50 $ eV). The soft
  X-ray excess emission is featureless and contributes $\sim 80\%$ to the total
  X-ray emission in the $0.3-10$ keV band. The most remarkable feature  of the
  present observation is the detection of X-ray variability at very  short time
  scale: the X-ray emission varied by 35\% in about 5000 s. We find that this
  variability can be explained by assuming that only  the overall normalization
  varied during the observation. There was no evidence for any short term spectral
  variability and the spectral shape was similar even during the {\it ASCA}
  observation carried out in 1997. Considering the high intrinsic luminosity
  ($\sim$ 2 $\times$ 10$^{45}$ erg s$^{-1}$) and the large inferred mass of the
  putative   black hole ($\sim1.6\times10^8 M_{\sun}$), the observed time scale of
  variability indicates emission at close to Eddington luminosity arising from
  very close to the black hole. We suggest that PHL~1092 in particular (and narrow
  line Seyfert galaxies in general) is a fast rotating black hole emitting close
  to its Eddington luminosity and the X-ray emission corresponds to the high-soft
  state seen in Galactic black hole sources.
\end{abstract}


\section{Introduction}
\par Narrow Line Seyfert 1 (NLS1) galaxies were first classified as a subclass of
Seyfert 1  by \citet{op85} on the basis that the FWHM of H$_\beta$ line is less
than 2000 km s$^{-1}$, the ratio of [O~III]$\lambda 5007$ to H$_\beta$ is less
than 3, and they often show strong Fe~II emission in $4500 - 4680$ {\rm \AA}
and $5105 - 5395$ {\rm \AA} regions. Though NLS1s are identified by their optical
properties they have even more remarkable properties in the X-ray band as compared
to other Seyfert galaxies. They show evidence for strong excess of soft X-rays
(dominant below $\sim$ 2 keV) above the hard X-ray continuum extrapolation and
rapid X-ray variability (doubling times of minutes to hours).

 \par Soft X-ray excess emission above a power-law continuum is usually identified
 as the steepening of the X-ray continuum below $\sim 2$ keV. This soft excess
 emission was first observed by {\it HEAO-1} \citep{pnnjwb81} and {\it EXOSAT}
 \citep{arn85,sgn85}. \citet{bbf96} showed that AGN with steepest soft X-ray spectra
 in the {\it ROSAT} band tend to lie at the lower end of the H$_\beta$ line width
 distribution and hence AGNs with soft excess emission are predominantly NLS1
 galaxies. Detailed X-ray study of NLS1s was carried out with {\it ASCA} which
 confirmed the soft X-ray emission and, in addition, showed that the $2-10$ keV
 continuum slope too is steeper and anti-correlated with the FWHM of the H$_\beta$
 line \citep{bme97,lei99b,vrwe99}. {\it ASCA} observations also showed that many
 NLS1s have Fe~K$_\alpha$ line  arising from highly ionized iron, and hence the
 accretion disks of NLS1s must be ionized \citep{dew02,prob04,bif01}. NLS1s are the
 most extreme X-ray variable objects among the radio-quiet AGNs. NLS1s frequently
 exhibit rapid (doubling time scale of a few hundred seconds) and/or  large (up to
 a factor of 100) amplitude X-ray variability \citep{bbf96,dsjmmn01,dbsl02}. The
 excess variance for NLS1s is typically an order of magnitude higher than that
 observed for Broad Line Seyfert 1's (BLS1).

 \par There are high-luminosity analogs of this class, the prototype of which is
 I~Zw~1 \citep{phi76}. I~Zw~1 also has weak forbidden lines, narrow permitted
 lines, and strong Fe~II lines. One of the most extreme narrow line Fe~II QSO is
 PHL~1092 (B=16.7, z=0.396) \citep{bk80,bk84,kcfzg95}. Its Fe~II$\lambda4570/H_\beta$
 ratio is 5.3, and its H$_\beta$ line width is  1800~km s$^{-1}$ \citep{bk84}.
 PHL~1092 was rapidly variable during a {\it ROSAT} observation, and its soft X-ray
 spectrum was extremely steep, modeled by a power-law with photon index more than
 4 \citep{fh96,bbfr99}. From an {\it ASCA} observation \citet{lei99b} found a huge
 soft excess (which can be described as a black body) in the PHL~1092 spectrum which
 is 80\% of the total flux. An evidence for a line at around 7 keV has also been
 found and this line could not be explained as an ionized iron line.

 \par  Recently \citet{dg04} made a detailed study of the XMM-Newton observations
 of I~Zw~1 and they showed that during the X-ray variability the power-law
 component varied without any change in the soft X-ray excess component. In
 contrast to this, PHL~1092 shows large variability and it has more than 80\%
 soft X-ray excess component. To understand the X-ray spectral variability of
 this high luminosity object we have analyzed  an {\it XMM-Newton} observation
 on this source. We present results based on this observation, including flux
 selected spectroscopy of high and low flux states.

 \begin{figure}
\includegraphics[angle=-90,scale=0.35]{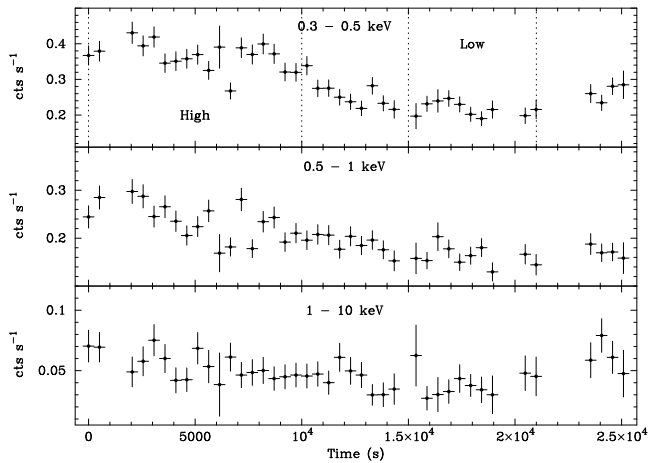}
 \caption{PN light curves of PHL~1092 in the 0.3 -- 0.5 keV
 (top panel), 0.5 -- 1 keV (middle panel) and 1 -- 10 keV
 (bottom panel) bands with a bin size of 512 s. The high and
 low flux states used for flux selected spectroscopy are
 shown by dotted vertical lines in the top panel of the figure.
 \label{f1}}
 \end{figure}

\section{Observation and Data reduction}
PHL~1092 was observed by {\it XMM-Newton} observatory on 2003 January 18 using
the European Photon Imaging Camera (EPIC) and the reflection grating spectrometer
(RGS). The raw events were processed and filtered using the most recent updated
calibration database and analysis software ({\tt SAS v6.0.0}) available in 2004
March. Events in the bad pixels and those adjacent pixels were discarded. Only
events with pattern $0 - 4$ (single and double) for the PN and $0 - 12$ for MOS
were selected. Examination of 16 s bin light curve extracted from the full field,
excluding the source region, and above 12 keV showed high particle background in
some observation intervals. The background flaring region is excluded using the
criterion 'rate $<3$' and minimum good time interval $>512$ s for PN data. This
resulted in 'good' exposure time of $\sim 18.9 {\rm~ks}$ for the PN. And during
these good time intervals the MOS1 and MOS2  background were reasonably low, so
we took the same time interval for MOS data also. The net count rate is {\bf
$0.53 {\rm~counts~s^{-1}}$} in PN and {\bf $0.12{\rm~counts~s^{-1}}$} in MOS1
and MOS2.

\section{Temporal Analysis}

\par X-ray light curves of PHL~1092 were extracted from the PN and MOS data
using a circular region of $35\arcsec$ and $45\arcsec$ respectively
centered at the source position and in the two soft bands ($0.3 - 0.5$ keV
and $0.5 - 1$ keV) as well as in the hard band ($1 - 10$ keV). The energy
bands were chosen to separate approximately the spectral components --  a
power-law (dominant at high energy) and a soft excess component generally
observed from NLS1 galaxies \citep[see][]{lei99b}. Background light
curves were extracted from source free regions with the same bin sizes,
exposure requirements, as for the source light curves. The area of the
region selected for the background is four times the area of the region
selected for the source. The background light curves were subtracted from
the respective source light curves after appropriate scaling to compensate
for the different areas of the extraction regions.

\par Figure~1 shows the PN light curves of PHL~1092 in the different
energy bands with a bin size of 512 s. It is evident that X-ray emission
from PHL~1092 varied strongly during the {\it XMM-Newton} observation
particularly in the soft bands ($ 0.3 - 0.5 $ keV and $ 0.5 - 1 $ keV).
There is a sharp decrease in the observed count rates starting from
10$^4$ s (from start of the observation) and reaches a low level in
about 5000 s. Two regions are marked as ``high'' (from start to 10$^4$ s)
and ``low'' (1.5 $\times$ 10$^4$ s to 2.1 $\times$ 10$^4$ s) in Figure~1
and demarcated by vertical dotted lines. A constant model fit to the
complete data gives an unacceptable value for the reduced $\chi^2$
($\sim 7.7$ and $\sim 3.7$ respectively in the two low energy bands).
Fitting constant models to the data in two regions (namely the high-flux
state and the low-flux state) gives much lower value of reduced $\chi^2
\sim 2.2$ and $\sim 0.73$ respectively in $ 0.3 - 0.5 $ keV band and
$\sim 2.9$ and $\sim 0.87$ respectively in the $0.5 - 1$ keV band.
The average count rate throughout the observation in the $ 0.3 - 0.5 $
keV band is $0.28 \pm 0.01$ s$^{-1}$, whereas in the high and low states
the average count rates are $0.36 \pm 0.01$ s$^{-1}$ and $0.22 \pm 0.01$
s$^{-1}$, respectively. In the $0.5 - 1 $ keV band the average count
rates in the high and low states are $0.23 \pm 0.01$ s $^{-1}$ and
$ 0.16 \pm 0.01 $ s$^{-1}$. During the whole observation the average count
rate was $0.19 \pm 0.01$ s$^{-1}$ in this energy band. This shows that
the observed count rates varied in a time scale of about 5000 s at a very
high significance level. Similarly the average count rate in the 1 - 10
keV band is $0.053 \pm 0.005$ s$^{-1}$ in high flux state and $0.036 \pm
0.006$ s$^{-1}$ in the low flux state, whereas the average count rate in
this band during the whole observation is $0.047 \pm 0.003$ s$^{-1}$.
So there is a clear variation of count rate in this band too.

\begin{figure}
\includegraphics[angle=-90,scale=0.35]{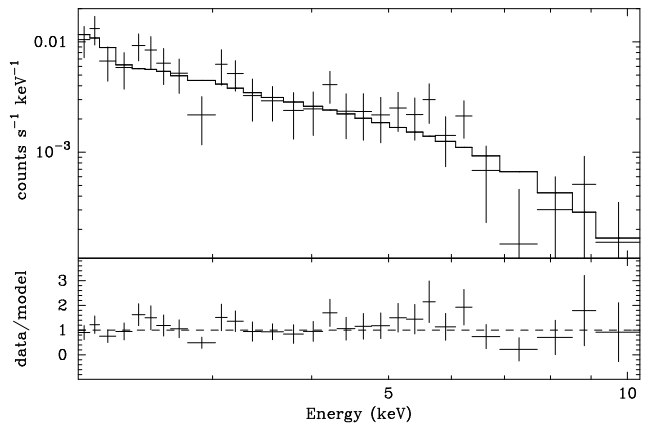}
\caption{2 -- 10 keV PN spectrum and best fitted model of absorbed and
red-shifted power-law.}
\label{f2}
\end{figure}

\section{Time-averaged spectral analysis}
\par Photon energy spectra of PHL~1092 and associated background spectra
were accumulated from the EPIC PN and MOS  data using the source and
background extraction method described above. We have also analyzed the RGS
data, but the observed count rates are too low to draw any definite
conclusion. The source spectra were grouped such that each bin contained
at least 15 counts. The EPIC responses were generated using the SAS tasks
{\tt rmfgen} and {\tt arfgen}. All the spectral fits  were performed with
the XSPEC package \citep[version 11.2.0;][]{arn96} and using the $\chi^2$
-statistic. Unless otherwise specified, the quoted errors on the best-fit
model parameters were calculated for the $90\%$ confidence level for one
interesting parameter i.e., $\Delta \chi^2 = 2.7$.

\subsection{2 - 10 keV spectrum}
\par We began studying the spectrum by fitting a red-shifted power-law model
with a Galactic $N_H$ of $3.53\times10^{20}$ \citep{mlle96} to the
EPIC spectrum in the energy range 2 - 10 keV. This provides an acceptable
fit giving a photon index $\Gamma\simeq1.88$ and a minimum $\chi^2(dof)
=22(27)$. The fit to the spectrum and the residuals are shown in Figure 2.
There is no clear indication of any line feature near 6 keV. The
non-detection of an Iron line is mainly due to the low level of detection
significance above 5 keV. The upper limit for the equivalent width of a
narrow iron line at 6.4 keV is 289 eV.

\begin{figure}
\includegraphics[angle=-90,scale=0.35]{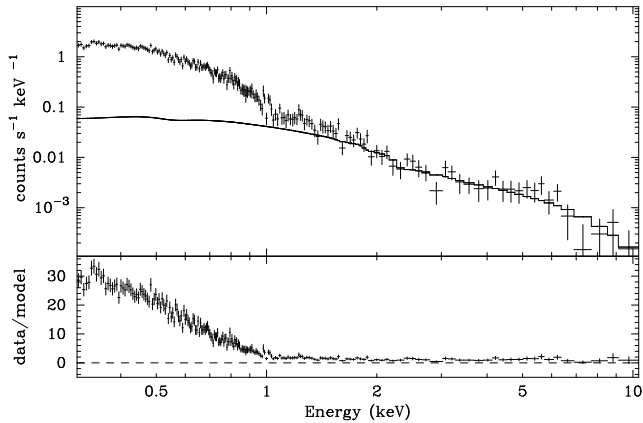}
\caption{The 0.3 -- 10 keV EPIC PN spectrum. The extrapolated power-law
model is also shown. A strong soft excess is present below $\sim$ 2 keV.}
\label{f3}
\end{figure}

\subsection{0.3 - 10 keV spectrum}
\par Extrapolation of the best fitting 2 - 10 keV power law to 0.3 keV shows
a huge soft excess in the spectrum (Figure~3). To fit the broad band
spectrum, we used one red-shifted blackbody component to parameterize the
soft excess. Adding a single blackbody provides a good fit $\chi^2\sim192(196)$
to the soft emission but the power law slope steepens to $2.15$. The blackbody
temperature is found to be $\sim124$ eV. Adding another blackbody gives an
improvement with $\Delta\chi^2\sim7$ for two extra parameters and the F-test
null probability for adding this component is 0.027. The temperatures of the
blackbody components are found to be 129 eV and 53 eV. The data and the
folded model are shown in Figure~4.  We have also tried using two red-shifted
power-law which gives high reduced $\chi^2$ and is unacceptable.

\begin{figure}
\includegraphics[scale=0.35,angle=-90]{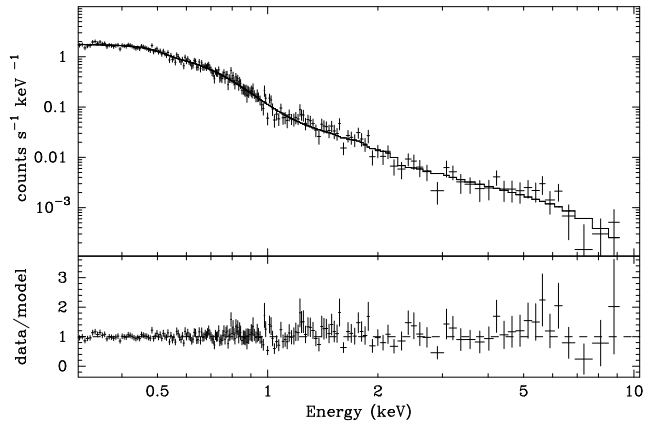}
\caption{PN spectral data and the best-fit model consisting of two absorbed
and red-shifted blackbody and a power-law ($\Gamma\sim2.1$).}
\label{f4}
\end{figure}

\subsection{PN and MOS data}

 We have made a joint fit to the PN and MOS data and find that the
 above model fits the data satisfactorily. The fit and the ratio of data to
 model are given in Figure 5 and the derived spectral parameters are given
 in Table 1. The ratio of data to model for the different instruments have
 been shown separately. In MOS1 there is a slight deficit of counts at around
 $\sim 1.4$ keV, whereas in the PN this feature is less significant and in
 MOS2 this is not present at all. Examining the data to model ratio one can
 argue that there is a slight excess emission at high energies. But we did
 not get any improvement in the fit by including line or edge parameters at
 the expected Fe lines and edges. To understand the nature of high energy
 emission properly we have put several markers (a,b,c,d,e,f) in Figure~5.
 Markers `a' and `b' are at 7.1 keV and 9.2 keV in the rest frame, respectively
 (corresponding to neutral and ionized edges of Fe). By examining the data
 to model ratio at these two energies one can conclude that there is no
 indication of edge in the data. Markers `c' and `d' correspond to 6.4 keV
 and 6.9 keV (corresponding to neutral and H-like Fe emission lines), and
 there is no apparent excess emission at these energies. The apparent line
 like emission feature in MOS2 data corresponds to unrealistically high energy
 (marker `e' at energy $\sim 8$ keV in the rest frame). Again if we consider
 the edge like feature at high energy in the

\begin{figure}[h]
\includegraphics[height=7.8cm,width=7.8cm]{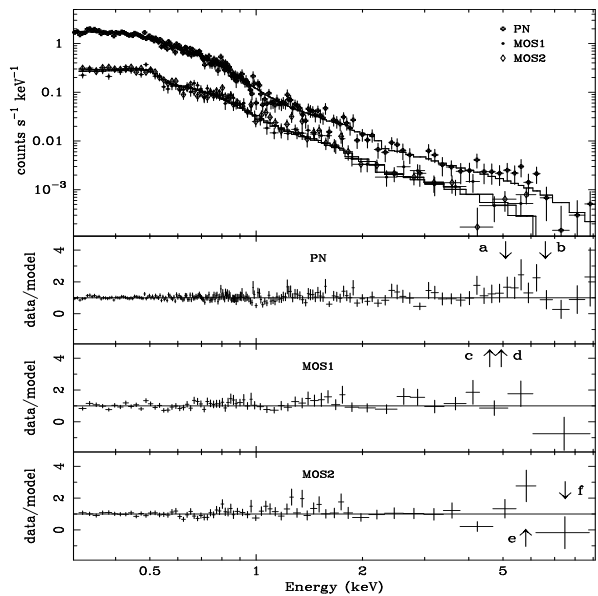}
\caption{PN, MOS1 and MOS2 spectral data and the best-fit model consisting
of two absorbed and red-shifted blackbody and power-law ($\Gamma\sim2.1$) (top
panel). Ratio of data to model for PN, MOS1, and MOS2 are also shown
separately for clarity. The labels a,b,c,d,e, and f correspond to the source
rest frame energies of 7.1 keV, 9.2 keV, 6.4 keV, 6.9 keV, 8 keV and 10.5 keV,
respectively (see text).}
\label{f5}
\end{figure}
\noindent
 data the corresponding energy will be $\sim 10.5$ keV (marker `f' in the
 figure) in the rest frame which is unrealistically high. Hence we can
 conclude that it is not possible to quantify the high energy excess by
 these known physical phenomenon. One possible reason of this excess emission
 at around 8 keV or deficit of counts at around 10.5 keV is an artifact of
 incorrect background subtraction. In this range background is dominated over
 the source counts giving rise to an apparent curvature. But if the curvature
 is true and due to some physical phenomenon then line emission from relativistic
 ionized accretion disk (laor) cannot be ruled out. By putting  a laor line
 we get an improvement of $\Delta\chi^2\sim8$ (F-test null probability $\sim$
 0.2), and the line energy is $\sim6.7$ keV in the rest frame (see section 6.4)
 for five extra parameters.

\begin{table}
\footnotesize
	\caption{\label{tab:tab1}Best-fit spectral parameters of PHL~1092.}
\begin{ruledtabular}
\begin{tabular}{lccccc}
& & \multicolumn{4}{c}{Model\footnotemark[1]} \\
	      & & \multicolumn{2}{c}{model 1\footnotemark[1]} & model
		  2\footnotemark[1]\\
Component & Parameter    & PN & PN+MOS &  PN \\
\hline
Blackbody & $kT$ (eV)  &  $129.0_{-4.4}^{+3.6}$ & $130.1_{-2.8}^{+3.4}$ & -- &\\
          & $n_{bb}\footnotemark[2]$  & $6.38_{-1.86}^{+0.65}$ & $6.04_{-0.60}^{+0.40}$  & --  & \\
          & $f_{bb}$ (0.3-10 keV) $\footnotemark[3]$ & $9.91 $  & $9.51$ &	-- & \\
Blackbody & $kT$ (eV)  &  $52.8_{-24.5}^{+13.3}$ & $54.3_{-6.5}^{+3.2}$ & -- & \\
          & $n_{bb}\footnotemark[3]$  & $13.0_{-11.2}^{+217.6}$ & $13.8_{-9.2}^{+87.0}$ & -- &  \\
          & $f_{bb}(0.3-10~keV)\footnotemark[4]$ & $1.12$  & $1.39$ & -- &  \\
Power law & $\Gamma$ &   $2.04_{-0.16}^{+0.21}$ & $2.15_{-0.15}^{+0.13}$ & -- & \\
          & $n_{pl} \footnotemark[4]$ & $1.19_{-0.19}^{+0.31}$ & $1.37_{-0.19}^{+0.18}$ & -- &\\
          & $f_{pl}$ (0.3-10 keV) $\footnotemark[3]$ & $2.95$ & $3.06$ & -- &  \\
Comptt	  &	$kT_{seed}$ (eV)		&	--	&	--	& $110_{-12}^{+10}$ \\
		  &	$kT_{plasma}$ (keV)	&	--	&	--	& $4.8_{-4.1}^{+12.0}$ \\
		  &	$\tau$ 					&	--	&	--	& $0.10_{-0.09}^{+0.39}$	\\
		  &	$n_C\footnotemark[5]$ 	&	--	&	--	& $1.4_{-0.9}^{+104.2}$	\\
		  &	$f_C$ (0.3 - 10 keV) $\footnotemark[3]$	&	--	&	--	& $11.40$	\\
Comptt	  &	$kT_{seed}$ (eV) 	&	--	&	--	& $0.32_{-0.32}^{+0.25}$ \\
		  &	$kT_{plasma}$ (keV)	&	--	&	--	& $59.2_{-57.2}^{+165.3}$ \\
		  &	$\tau$ 					&	--	&	--	& $0.51_{0.50}^{+5.05}$ \\
		  &	$n_C\footnotemark[5]$ 	&	--	&	--	& $0.002_{-0.001}^{+0.870}$ \\
		  &	$f_C$ (0.3 - 10 keV) $\footnotemark[3]$ &	--	&	--	& $2.26$ \\
		  \\
Total     & $f_{obs}$ (0.3-10 keV) $\footnotemark[3]$ & $14.0$ & $14.0$ & $13.6$ \\
          & $f_{intr}$ (0.3-10 keV) $\footnotemark[3]$ & $20.8$ & $20.9$ & $19.9$ \\
	      & $L_{intr}$ (0.3-10 keV) $\footnotemark[6]$ & $20.1$ & $20.2$ & $19.2$ \\
          & $\chi^2_{min}/dof$ & $186/194$ & $334/326$ & $189/191$ \\
\end{tabular}
\end{ruledtabular}
\footnotetext[1]{Parameter values for spectral fits EPIC PN and MOS data.
Spectral models were modified by the Galactic absorption. ($N_H^{Gal}=3.53
\times10^{20}{\rm~cm^{-2}}$). model 1: WABS(BB+BB+PO), model 2: WABS(COMPTT
+COMPTT) where WABS: photoelectric absorption, BB: blackbody PO: power-law,
COMPTT: Comptonization}
\footnotetext[2]{Blackbody normalization in units of $10^{-5}\times10^{39}
{\rm~erg~s^{-1} / (d_{10})^2}$, \\ where $d$ is the distance in the unit of
10 kpc}
\footnotetext[3]{Flux in the unit of $10^{-13}{\rm~erg~cm^{-2}~s^{-1}}$}
\footnotetext[4]{Powerlaw normalization in units of
$10^{-4}{\rm~photons~cm^{-2}~s^{-1}} \\ {\rm keV^{-1}}$ at $1$ keV}
\footnotetext[5]{Comptonization normalization in units of $10^{-3} {\rm
photons~cm^{-2}} \\ {\rm s^{-1}~keV^{-1}}$}
\footnotetext[6]{Source luminosity in the unit of $10^{44}{\rm~erg~s^{-1}}$}
\end{table}

\subsection{Two component thermal Comptonization model}
The red-shifted and absorbed blackbodies and power-law model described
the data adequately,  but the temperature of the blackbodies ($kT \sim
130$ eV) and ($kT \sim 50 $ eV) are unrealistically high for a standard
accretion disk (see section 6.3). A corona with two electron populations
with distinct temperatures can give rise to both a soft X-ray excess and
a power-law. We investigated if such a process is a viable model for the
X-ray spectrum of PHL~1092. The contribution of this soft excess component
to the $0.3-2$ keV band unabsorbed emission is $\sim 80\%$. As the
temperature of the soft excess component is much higher than that expected
from an accretion disk, this component is unlikely to be the intrinsic
disk emission. Replacing the soft component by  {\tt comptt}  we get a
good fit ($\chi^2 = 190$ for 193 dof). Then we replace the  power-law
with a hotter comptonized component. The fit is as good as  previous
($\chi^2 = 189$ for  193 dof), since the EPIC instrument cannot determine
the value of the electron temperature if $kT\gtrsim4$ keV. The exponential
roll-over of the comptonized component occurs at $\sim 4kT$, and below
this point the emission appears as a power-law. Thus if $4kT$ is very much
greater than $\sim 10$ keV, the temperature cannot be constrained over
the XMM-Newton band, and the fit is indistinguishable from a power-law
over the same  energy range. Because of this the resultant $\chi^2$ value
for the high temperature Comptonization fit is virtually identical with
the power-law fit. The parameter values of two comptonization fit are
given in Table~1.

\section{Flux-selected spectroscopy}
\par To investigate whether the flux variability is due to changes of the
spectral parameters we have carried out a spectral analysis at low and high
flux levels (shown in Figure~1 by dotted vertical lines). We extracted two
average spectra covering 0 - 10 ks (high flux state) and 15 - 21 ks (low
flux state) in the 0.3 - 10 keV band. The average PN count rates are $0.64
\pm 0.02$ counts s$^{-1}$ (high flux state) and $0.41 \pm 0.02$ counts s$^{-1}$
(low flux state). We have used the model (absorbed and red-shifted power-law,
and two blackbody) for both the spectra. First we fixed all the parameter
previously found values for the combined fit of PN and MOS data and obtained
a value of $\chi^2$(dof) of 242(141) and 285(85) for the high and low flux
data, respectively. These values improved to 165 and 77 when the relative
normalization of the two data sets are changed. The fit improved marginally
($\chi^2$(dof) of 152(144) and 76(87), respectively) when all the
normalizations are allowed to vary mutually. The observed count rates
are too low to investigate spectral variability at a greater detail. In
Table 2 the fit parameters and the observed fluxes are given. It can be seen
from the table that the source is highly variable during the observation.
The change in luminosity is $\sim35\%$ between the two states.

\begin{table}
\footnotesize
\caption{\label{tab:tab2}Best-fit spectral model parameters derived from
the spectral fitting of the PN spectra of PHL~1092.}
	\begin{ruledtabular}
\begin{tabular}{lcccc}
	  &          & \multicolumn{2}{c}{Flux States\footnotemark[1]} \\
Component & Parameter & (high) & (low) \\
\hline
\\
BB (kT$\sim$130 eV) & $n_{bb}\footnotemark[2]$ & $6.87_{-0.27}^{+0.28}$ & $4.59_{-0.28}^{+0.43}$ \\

BB (kT$\sim$50 eV)  & $n_{bb}\footnotemark[2]$ & $24.6_{-6.6}^{+6.7}$ & $8.9_{-3.7}^{+8.3}$ \\

PO & $n_{pl}\footnotemark[3]$  & $1.33_{-0.13}^{+0.13}$ & $0.74_{-0.55}^{+0.74}$ \\
\\
Total	  & $f_{obs} $ (0.3-10 keV) $\footnotemark[4]$ & $16.01$ & $10.07$ \\
          & $f_{intr}$ (0.3-10 keV) $\footnotemark[4]$ & $24.53$ & $14.91$ \\
	      & $L_{ints}$ (0.3-10 keV) $\footnotemark[5]$ & $23.64$ & $14.37$ \\
          & $\chi^2_{min}/dof$ & $152/144$ & $75/88$ \\
\\
\end{tabular}
\end{ruledtabular}
\footnotetext[1]{High: High flux state; Low: Low flux state}
\footnotetext[2]{Blackbody normalization in units of $10^{-5}
\times 10^{39}{\rm~erg~s^{-1}/(d_{10})^2}$, where $d$ is the
in 10 kpc distance.}
\footnotetext[3]{Power-law normalization in units of $10^{-4}
{\rm~photons~cm^{-2}~s^{-1}} \\ {\rm keV^{-1}}$ at 1 keV.}
\footnotetext[4]{Flux in units of $10^{-13}{\rm~erg~cm^{-2}~s^{-1}}$}
\footnotetext[5]{Source luminosity in units of $10^{44}{\rm~erg~s^{-1}}$}
\end{table}

\section{Discussion}

This paper presents an analysis of $\sim 20{\rm~ks}$ {\it XMM-Newton}
observation of the narrow-line  QSO PHL~1092. The time-averaged X-ray spectrum
of PHL~1092 shows a power-law continuum, a soft X-ray excess described by two
red-shifted blackbody components below $\sim 2$ keV. Above $2.0$ keV, the
power-law component has a slope of $\Gamma \sim 2.1$ which is steeper than
that of most BLS1 galaxies ($\Gamma \sim 1.9$) but is similar to other NLS1
galaxies e.g, Ton~S180 \citep[$\Gamma=2.26_{-0.12}^{+0.05}$;][]{vbfbbt02},
Mrk~335 \citep[$\Gamma=2.29 \pm 0.02$;][]{gols02}, Akn~564 \citep[$\Gamma\simeq2.50
- 2.55$;][]{vbbfv03}. The X-ray continuum steepens at lower energies, resembling
the soft X-ray excess emission generally observed from NLS1 galaxies. The
strength of this soft X-ray emission is quite high in PHL~1092 compared to other
NLS1s. The contribution of the soft X-ray excess emission to the $0.3 - 2$ keV
band X-ray emission is about $80\%$ in PHL~1092. Such high soft excess has been
reported earlier by \citet{lei99b}  based on the ASCA data. Variability at
similar variability time scale has been reported by \citet{fh96} based on the
ROSAT data.

\subsection{Black hole mass for PHL~1092}
Before we discuss the results in detail, we estimate the mass of the black hole
for PHL~1092. Calculations of the black hole mass rely on the assumption that the
dynamics of the optical broad-line region gas is dominated by the gravitational
potential of the central super-massive black hole. To calculate the black hole
mass for PHL~1092, we used the results of Kaspi et al. (2000), based on a
reverberation study of 17 quasars. \citet{ksnmjg00} have determined an empirical
relationship between the size of the broad-line region ($R_{BLR}$) and the
monochromatic luminosity ($\lambda L_{\lambda}$) as follows.

\begin{equation}
	R_{BLR} = (32.9_{-1.9}^{+2.0}) \left[\frac{\lambda L_{\lambda}(5100)}
	{10^{44} {\rm~erg~s^{-1}}}\right]^{0.700 \pm 0.033} {\rm~lt-days}
\end{equation}
We calculated the flux approximately from the IDS spectrum of PHL~1092 obtained
from \citet{bk80} at the rest wavelength of 5100 {\rm~\AA} ($\lambda f_{\lambda}$
(5100 {\rm~\AA}) $\sim 3\times10^{-12}$ ${\rm~erg~cm^{-2}~s^{-1}}$). The mass of
the black hole is given by $M_{BH}=rv^{2}G^{-1}$. To determine $v$, the velocity,
we correct $v_{FWHM}$ of the H$_\beta$ emission line by a factor of $\sqrt{3}/2$
to account for velocities in three dimensions. We used $v_{FWHM} = 1790$ km
s$^{-1}$ for the H$_\beta$ line from \citet{lei99a}. The mass is then,

\begin{equation}
M = 1.464 \times10^{5}\left(\frac{R_{BLR}}{\rm~lt-days}\right)
\left(\frac{v_{FWHM}}{10^3{\rm km~s^{-1}}}\right)^{2} {\rm~M\odot}
\end{equation}

Using the cosmology $q_{0} =0$ and $H_{0}=50$ km s$^{-1}$Mpc$^{-1}$, this gives
us a value for the central black hole mass of $\sim (1.6 \pm 0.6) \times 10^8
M_{\sun}$. \citet{cnpk01} have reported that the mass of black hole in PHL~1092
is $1.2\times10^6 M_{\sun}$ from X-ray variability studies  and it is $1.8 \times
10^8 M_{\sun}$ from accretion disk fitting.

\subsection{Rapid variability and the efficiency limit}
PHL~1092 showed rapid and large amplitude X-ray variability during the {\it
XMM-Newton} observation. X-ray emission from PHL~1092 decreased by $\sim35\%$
in $5000$ s, corresponding to a change in the $0.3 - 10$ keV luminosity of
$\Delta L \sim 10^{45}{\rm~erg~s^{-1}}$. This is a remarkable variability for
a radio-quiet quasar as the time scale is a factor of $\sim 10$ shorter than
the dynamical time scale for a $10^8 M_{\sun}$ Schwarzschild black hole
but comparable to the dynamical time scale of Kerr black hole of same mass.
Such variability is rare among radio-quiet quasars. Other radio-quiet quasars
to show such rapid and large amplitude variability are I~Zw~1 \citep{dg04},
RX J1334.2+3759 \citep{dbsl02}.

\par The major consequence of rapid and large amplitude variability is that
the source has to emit very efficiently. It is possible to calculate the
efficiency of conversion of rest mass into energy using simple arguments
originally due to \citet{fab79}. If the X-ray emitting region is an accreting
spherical cloud of ionized hydrogen, then there is a limit on the radiative
efficiency, $\eta$, given by $\eta > 4.8\times 10^{-43} \Delta L/\Delta t$,
where $ \Delta L$ is the change in luminosity in the rest-frame time interval
$\Delta t$. The change in luminosity between the two states is $\sim5 \times
10^{44}$ ergs s$^{-1}$ in 5000 s. Assuming uniform and spherical emission this
gives a lower limit to the efficiency, $\eta \gtrsim 0.1$. This efficiency
can be compared with the maximum efficiency possible for a Schwarzschild black
hole \citep[$\eta \simeq 0.06$;][]{st83} and a Kerr black hole \citep[$\eta
\simeq 0.3$;][]{tho74}. Thus the rapid variability observed from PHL~1092 is
consistent with that expected for a Kerr black hole.

\subsection{The soft X-ray excess emission}
Although the soft X-ray excess emission above a power-law continuum is well
described by a blackbody component, its temperature is too high for the
intrinsic emission from an accretion disk. There is a relationship between
the disk temperature of a standard thin accretion disk, accretion rate and
the mass of the black hole, given by \citep{pet97},

\begin{equation}
T(r) \sim 6.3\times 10^5 \dot{m}^{1/4} M_8^{-1/4}
\left (\frac{r}{r_S} \right)^{-3/4} {\rm~ K}
\end{equation}

where $r_S$ is the Schwarzschild radius, and $M_8$ is the black hole
mass in units of $10^8 M_{\sun}$. The temperature at the last stable
orbit ($r=3r_S$) for a black hole is $\sim 22~eV$ for a black hole
mass of $1.6 \times 10^8 M_{\sun}$ accreting at the Eddington rate. This
is an upper limit to the disk temperature, because $\dot{m} < 1$, and
outer regions of the disk will have lower temperatures. This can be
reconciled by assuming that the low energy excess is due to a population
of electrons very close to the black hole giving a Comptonized spectrum.
In fact the observed X-ray spectrum in the 0.3 - 10 keV region can be
fit with a two component Comptonization model. An alternative viable
solution is the 'slim disk' \citep[][and references therein]{acls88,cw04}
which allows temperature higher than the standard disk, and takes place
at high accretion rates.

\begin{figure}
\includegraphics[height=7.8cm,width=7.8cm]{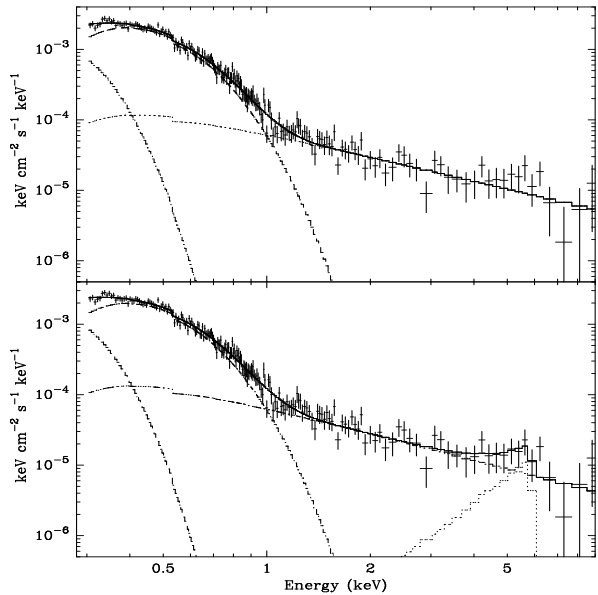}
\caption{Unfolded PN spectral data and best fit model components: two absorbed
and red-shifted blackbody and power-law (above) and two absorbed and
red-shifted blackbody, power-law and a laor line (below).
\label{f6}}
\end{figure}

\subsection{Possible evidence for a high-energy curvature and absorption}
Recently \citet{gbbfg04} reported a high energy curvature around $\sim 7$
keV in the same data of PHL~1092. They interpreted the high energy curvature
as due to light bending effects very close to the black hole. When we
analyze the data using the latest analysis tools (SAS 6.0.0) and latest
calibration, we find very little evidence for a high energy curvature and
the curvature is centered around 7.5 keV. A narrow Gaussian line cannot
give satisfactory fit to the high energy curvature. However a broad Gaussian
(width $\sim 1 $ keV) can give an visual improvement of the spectrum (but
not statistically significant, $\Delta\chi^2\sim3$ for three extra
parameters, F-test null probability $\sim$ 0.4) but the line energy becomes
greater than $7.2 $ keV which is clearly unphysical. However a line from a
disk around a rotating black hole \citep{lao90} can be used to handle the
high energy curvature. The fit is marginally better than the previous
fit (see section 4.3). The unfolded spectrum without the Laor line is
given in the top panel of Figure~6 and with the line is given in the
bottom panel. The line energy is $\sim 6.7$ keV and the equivalent width
is $\sim 2.5 $ keV. Hence we can conclude that Gaussian lines are not
required to fit the data, but Laor lines from ionized accretion disk
cannot be ruled out. The large equivalent width of the line seems
unphysical and in contrast with the fact that the line equivalent width
diminishes with increasing continuum luminosity (Baldwin effect).
\citet{gbbfg04} also reported an absorption line around $\sim 1.4 $ keV.
This absorption feature was not present during the 1997 ASCA observation
(Leighly 1999b). In section 4.3 we discussed why inclusion of an absorption
line or an edge is not feasible. The major reason for the difference in
the results given in Gallo et al. (2004) and the present work is the use of
most recent calibration and SAS version.

\begin{figure}
\includegraphics[height=7cm,width=7.8cm]{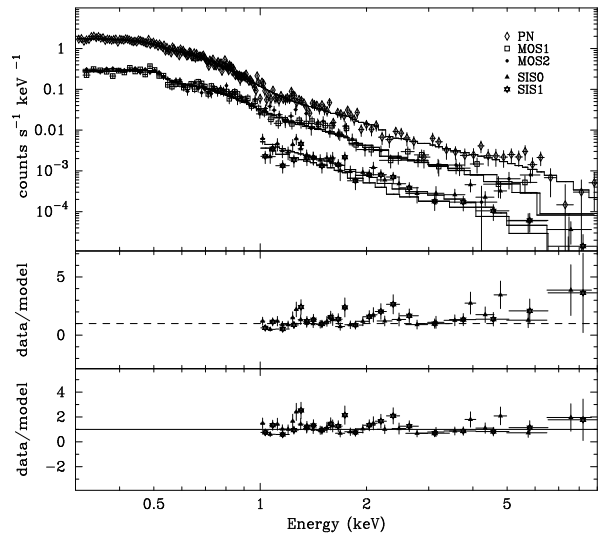}
\caption{PN, MOS1, MOS2, ASCA-SIS0, ASCA-SIS1 spectral data and the
best-fit model consisting of two absorbed and red-shifted blackbody
and power-law (top panel). ASCA-SIS data to model ratio freezing the
power-law index of ASCA data to that found in XMM-Newton data (middle
panel) and the same model but keeping the value of power-law index of
ASCA data free (lower panel). PN, MOS1 and MOS2 data to model ratio
have not been shown for clarity.}
\label{f7}
\end{figure}

\par To further understand the high energy spectral shape, we have
reanalyzed the ASCA data observed on 1997 July \citep{lei99a,lei99b}.
The ASCA-SIS0 and ASCA-SIS1 data are simultaneously fitted with the XMM-Newton
EPIC-PN and EPIC-MOS data. To avoid the calibration uncertainties of
ASCA-SIS (see ASCA calibration page) below 1 keV we take 1 - 10 keV energy
band. We fixed all the parameters obtained from the joint  fit of EPIC-PN
and EPIC-MOS. The  relative normalizations between the  instruments are
allowed to vary. The relative normalizations of ASCA-SIS0 and ASCA-SIS1 are
1.51 and 1.38 respectively. The fit obtained is quite satisfactory
($\chi^2\sim397$ for 377 dof), indicating that over the different epochs of
observation variability can be explained by overall flux variation rather
than spectral parameter variation. The observed and fitted spectra are shown
in Figure~7, along with residuals  (plotted as ratio of the data to the
model in the bottom panel of the figure). It can be seen that there is a
reasonable agreement between the data and the model. Changing the power-law
index improved the fit with a $\Delta \chi^2$ of 17, with the power-law index
changing to 1.73.  In this case the high energy excess is even lower (see
Figure~7 bottom panel). We can conclude that once the low-energy spectral
shape is appropriately modeled, there is no necessity of invoking any high
energy curvature. The data quality, however, is insufficient (particularly
above 5 keV) to make any definitive statements about other possible
spectral features like edges, reflection features etc.

\subsection{Comparison with the high-soft state of Galactic black hole sources}

 All the available data suggest a smooth spectrum with all the spectral
 features varying at a fast time scale. The time scale of variability,
 however, corresponds to a physical size (light crossing time) of 1.5
 $\times$ 10$^9$ km, which corresponds to only 3 Schwarzschild radius for
 a 1.6 $\times$ 10$^8$ M$_\odot$ black hole. Since both the spectral
 components appear to be varying in this time scale, we can conclude that
 the X-ray emission arises from very close to a rotating black hole. Such
 emission mechanism is seen in the high-soft states of Galactic black hole
 (GBH) binary sources. The accretion disk reaches very close to the black
 hole and bulk motion near to the black hole can give rise to the observed
 rapid  variability \citep{mr04}. The rapid variability observed in PHL~1092
 on time scale $\la$ 10000 s can be compared with the millisecond flares
 observed in GBHs \citep[e.g.,][]{zg04}. \citet{mpupm04} have compared the
 variability of NGC~4051 in long  and short time scales with that of Cygnus
 X-1 and have concluded that the X-ray  variability in NLS1 galaxies scale
 better with Cyg X-1 in its high state.

 \par The observed 0.3 - 10 keV X-ray luminosity corresponds to $7\%$ (or
 $10\%$ after correcting for Galactic absorption) of the Eddington
 luminosity ($L_{Edd}$). If X-ray luminosity represents a fraction
 (typically $\sim 10\%$) of  bolometric luminosity then the bolometric
 luminosity of PHL~1092 will be $\sim$  L$_{Edd}$. The high-soft states of
 Galactic black hole sources show a total X-ray  luminosity above $\sim$
 0.1 L$_{Edd}$ \citep{zg04}.

\par For the fast correlated X-ray variability, we can have the following physical
picture. The low energy blackbody arises from a optically thick disk extending
up to the last stable orbit of a rotating black hole. Within this radius, matter
gets accreted at a supersonic velocity which can produce bulk-motion Comptonisation
manifested as the power-law. Non-thermal activity (by the possible magnetic field
generated by the equipartition of energy) can generate high energy electrons which
would be cooled by Comptonization by the copious amount of low energy photons
from the low-energy blackbody emission. This can produce highly saturated Comptonization
spectrum modeled as a blackbody in the present work. We invoke the bulk-motion
Comptonization model given by \citet{ct95} to estimate the relevant energetics for
such a simplified model. The total luminosity of the $\sim$ 50 eV blackbody
corresponds to $\sim 5.4\times10^{45}~ergs~s^{-1}$ which is the expected luminosity
from a thick accretion disk extending up to 3 Schwartzchild radius for a 1.6 $\times$
10$^8$ M$_\sun$ rotating black hole with 0.2 Eddington accretion rate. The power-law
luminosity is about 10\% of the total luminosity, which is similar to that seen
in Galactic black hole sources in their high state \citep{zg04}. Comparison of
the timing and spectral properties of such high luminous NLS1 galaxies with the
high soft states of Galactic black hole sources would provide additional handle
to constrain such a model.

\section{Conclusions}
We presented spectral characteristics of the narrow line  QSO PHL~1092
based on an {\it XMM-Newton} observation. The main results are as follows.
\begin{enumerate}
\item The $0.3-10$ keV spectrum of PHL~1092, obtained with XMM-Newton,
	consists of three intrinsic spectral components, namely a steep
	($\Gamma_X \sim 2.1$) primary continuum described by a power-law,
	a soft X-ray excess component described by two red-shifted blackbody
	($kT \sim 130$ eV and $kT \sim 50 $ eV). Iron fluorescent lines are
	not detected, primarily due to the low detection level at $>$ 6 keV.
    The intrinsic photoelectric absorption component is consistent with
	the Galactic absorption.
\item PHL~1092 showed a significant X-ray variability with changes in the
	luminosity $\Delta L \sim 10^{45}{\rm~erg~s^{-1}}$ on a time scale of
	$\sim 5000$ s. The radiative efficiency, $\eta \gtrsim 0.1$, inferred
	from the variability is consistent with X-ray emission outside the last
	stable orbit around a Kerr black hole.
\item The long (comparing {\it ASCA} and {\it XMM-Newton} observation) and
	short term variability of PHL~1092 can be explained by change in
	overall normalisation without any spectral changes.
\item The existing data do not require any high energy curvature, but we
	cannot completely rule out the existence of a red-shifted line from the
	accretion disk near a fast rotating black hole (Laor line).
\end{enumerate}

\acknowledgements This work is based on observations obtained with {\it
	XMM-Newton}, an ESA science mission with instruments and contributions
	directly funded by ESA Member States and the USA (NASA). SD would
	like to acknowledge the partial support from the Kanwal Rekhi Scholarship
	of the TIFR Endowment Fund. We thank an anonymous referee for useful
	comments and suggestions.

\end{document}